\documentclass[epjST]{svjour}
\usepackage{graphics}
\usepackage{graphicx}
\usepackage{amssymb}
\usepackage{lineno}
\usepackage{bm}
\usepackage{pifont}
\usepackage{color}
\usepackage{marvosym}
\usepackage[outercaption,% caption placed on the right side
	]{sidecap}
\usepackage{pifont}
\usepackage{wrapfig}
\usepackage{amsfonts}
\usepackage{slashed}
\renewcommand\sidecaptionsep{1em}
\sidecaptionvpos{table}{c}
\sidecaptionvpos{figure}{c}

\begin{document}
%\sidecaptionsep{4cm}
%
\title{Hypernuclei and in-medium chiral dynamics}
%\subtitle{($\dots$on the microscopic origin of the spin-orbit interaction)}
\author{Paolo
 Finelli\inst{}\thanks{\email{paolo.finelli@bo.infn.it}}}
% \and Second author\inst{2} \and ... }
%
\institute{Physics Department, University of Bologna,\\Via Irnerio 46, 40126 Bologna (Italy)}
%\and the second here \and ...}
%
\abstract{
A recently introduced relativistic nuclear energy density functional, 
constrained by features of low-energy QCD, is extended 
to describe the structure of hypernuclei.
The density-dependent mean field and the spin-orbit potential 
of a $\Lambda$-hyperon in a nucleus,
are consistently calculated using the $SU(3)$ extension of in-medium 
chiral perturbation theory.
The leading long-range $\Lambda N$ interaction arises from
kaon-exchange and $2\pi$-exchange with a $\Sigma$-hyperon
in the intermediate state.
Scalar and vector mean fields, originating from in-medium changes 
of the quark condensates, produce
a sizeable {\em short-range} spin-orbit interaction.
The model, when applied to oxygen as a test case,
provides a natural explanation for the 
smallness of the effective $\Lambda$ spin-orbit potential: an almost complete
cancellation between the background contributions (scalar and vector) and 
the long-range terms generated by two-pion exchange.
} %end of abstract
\maketitle
%

%%%%%%%%%%%%%%%%%%%%%%%%%%%%%%%%%%%%%%%%%%%%%%%%%%%%%%%%%%%
%%%%%%%%%%%%%%%%%%%%%%%%%%%%%%%%%%%%%%%%%%%%%%%%%%%%%%%%%%%

%Brief introduction

\section{Hypernuclear phenomenology } 
%Basic features of hypernuclei

\begin{figure}[t]
\begin{center}
\includegraphics[scale=0.40]{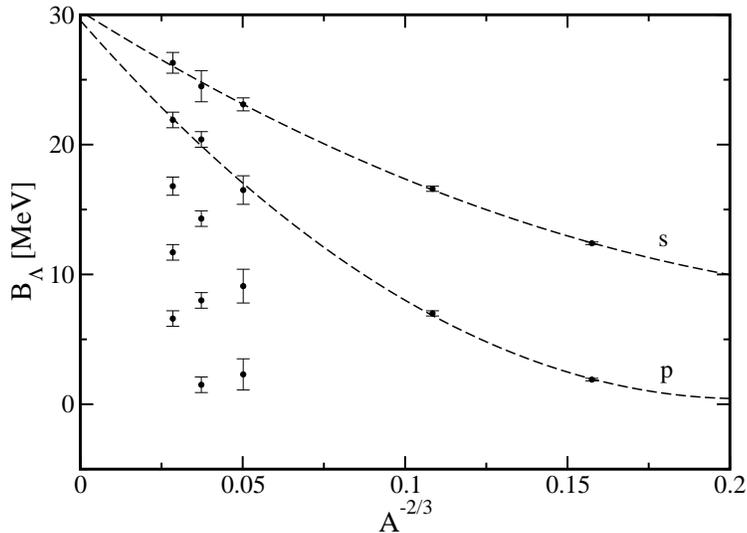}
\caption{\label{FigA}
Data on binding energies $B_\Lambda$ of $\Lambda$ single-particle
states (in various orbits) as functions of $A^{-2/3}$, where A is the mass
number of the nuclear core (see also Ref.~\cite{Millener:1988hp}).
The experimental values have been taken 
from Ref.~\cite{Hashimoto:2006aw} (Tables 11 and 13).
The dashed curves extrapolate the values of $B_\Lambda$
to the nuclear-matter limit ($A^{-2/3} \rightarrow 0$).
}
\end{center}
\end{figure}

\begin{figure}[h]
%\hspace{-1cm}
\begin{center}
\includegraphics[scale=0.45,angle=0]{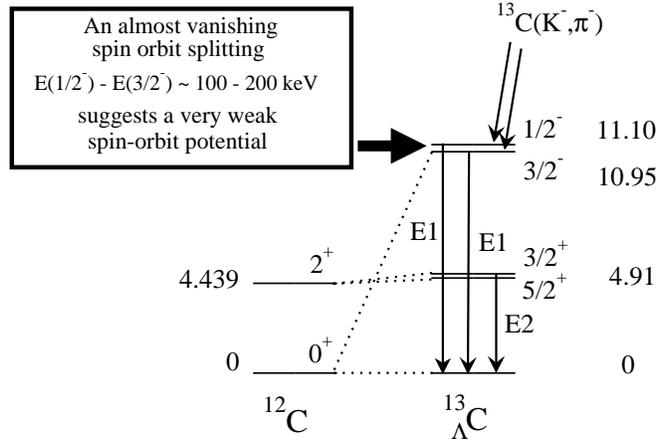}
%\end{center}
\caption{\label{FigB}
The hypernuclear $\gamma$ transitions observed in 
$^{13}_{\Lambda}{\rm C}$~\cite{Hashimoto:2006aw,Kohri:2001nc}.
The levels are denoted by their excitation energy (in MeV), and the 
spin and parity. The observed spin-orbit splitting is small
($154 \pm 54 \pm 36$ keV), i.e. $20\sim 30$ times smaller than that 
of the nucleon {\it p}-levels.
The ordering of the $\Lambda$ {\it p}-levels appears to be the same 
as for the nucleon levels. 
(From Fig.~47 in Ref.~\cite{Hashimoto:2006aw}).
}
\end{center}
\end{figure}

A hypernucleus is a nucleus in which one or more nucleons have been 
replaced by strange baryons~\cite{danysz,chriendover}.
In particular, $\Lambda$-hypernuclei are quantum systems composed of 
a single $\Lambda$-hyperon plus a core of nucleons.
The standard notation, used throughout this work, is 
\[
~^{A}_{\Lambda}{\rm Z} \quad {\rm where} \quad \left\{
\begin{array}{l}
A:~{\rm ~total~number~of~baryons~(nucleons~+~hyperon)} \\
{\rm Z}:~{\rm ~total~charge~(not~necessarily~the~number~of~protons)}\\
\Lambda:~{\rm ~hyperon~(in~this~case,~but~in~general,~also~\Sigma, \Xi,\ldots)}
\end{array}
\right. 
\]
For instance, $~^{13}_{\Lambda}{\rm C} = 6p + 6n + \Lambda$.
A simple way to describe hypernuclear properties is to employ
a phenomenological potential~\cite{Bando:1990yi}
\begin{equation}
U^\Lambda(r) = U_c^\Lambda(r) + U_{ls}^\Lambda(r) \;,
\end{equation}
where 
\begin{equation} 
U_c^\Lambda (r) =  -V_c^\Lambda f(r) \quad \quad {\rm and} \quad \quad
U_{ls}^\Lambda = V_{ls}^\Lambda \left(\frac{\hbar}{m_\pi c}\right)^2
\frac{1}{r} \frac{df(r)}{dr} {\bm s} \cdot {\bm l} \; , 
\end{equation}
are the central and spin-orbit terms,  respectively. 
$f(r)$ is a radial form factor that can be determined 
from the corresponding nuclear density distribution 
$f(r)= \rho(r)/\rho(0)$, or chosen in a Woods-Saxon form
$f(r) = 1 / \{1 + exp[(r-R)/a] \}$. 
In Fig.~\ref{FigA} we display the $\Lambda$ binding energies $B_\Lambda$
for a set of hypernuclei,  as functions of $A^{-3/2}$.
The dashed curves are extrapolations to the nuclear-matter 
limit $(A \rightarrow \infty)$. When compared to the binding of a nucleon, 
one notes a considerable reduction of $B_\Lambda$. 
The strength of the central potential $V_c^\Lambda$ is estimated~\cite{Bando:1990yi}\footnote{
This analysis is based on data for light and medium hypernuclei
(from$~^{12}_{\Lambda}{\rm C}$ to$~^{40}_{\Lambda}{\rm Ca}$), 
but the inclusion of heavier hypernuclei
does not change the estimate of Eq.~(\ref{central}) significantly 
(28 MeV for a fit with a Woods-Saxon potential~\cite{Millener:1988hp}).} 

\begin{equation}
V_c^\Lambda \simeq 32 \pm 2~{\rm MeV} \;,
\label{central}
\end{equation}
i.e. about $2/3$ of the depth of the nucleon potential. 
On the other hand, the $\Lambda$-nucleus spin-orbit interaction
is extremely weak~\cite{Bando:1990yi}:
\begin{equation}
V_{ls}^\Lambda \simeq 4 \pm 2~{\rm MeV} \;.
\label{spinorbit}
\end{equation}
This peculiar property has  recently been  confirmed by the E929 BNL experiment
\cite{Kohri:2001nc}, which observed
the spin-orbit splitting between $(p_{1/2})_\Lambda$ and $(p_{3/2})_\Lambda$
orbits in$~^{13}_{\Lambda}{\rm C}$, by separately detecting the $\Lambda$ inter-shell
transitions $(p_{1/2})_\Lambda \rightarrow (s_{1/2})_\Lambda$
and $(p_{3/2})_\Lambda \rightarrow (s_{1/2})_\Lambda$ around 11 MeV (see Fig.~\ref{FigB}).
The observed energy spacing is $E(1/2^-) - E(3/2^-) = 152 \pm 54 \pm 36$ keV.
Measurements of such high precision are feasible because the spreading widths of the 
$\Lambda$ hypernuclear states are extremely small~\cite{Hashimoto:2006aw}.

Considering that the corresponding spin-orbit strength for the nucleon
is an order of magnitude larger ($V_{ls}^N \sim 20$ MeV), it is extremely important
to understand the microscopic mechanism at the basis of this unnatural
property.

%%%%%%%%%%%%%%%%%%%%%%%%%%%%%%%%%%%%%%%%%%%%%%%%%%%%%%%%%%%
%%%%%%%%%%%%%%%%%%%%%%%%%%%%%%%%%%%%%%%%%%%%%%%%%%%%%%%%%%%

\section{In-medium chiral dynamics applied to hypernuclei}
%Brief introduction
\label{sec2}

A novel approach to the nuclear many-body problem, 
based on in-medium chiral effective field theory, has recently been 
succesfully applied to nuclear matter and properties of finite nuclei 
\cite{Kaiser:2001jx,Fritsch:2004nx,Finelli:2003fk,Finelli:2005ni}. In particular, 
it has been demonstrated that iterated one-pion exchange and irreducible $2\pi$ exchange processes, with inclusion of Pauli blocking effects and $\Delta$-isobar 
excitations in intermediate states, generate the correct nuclear binding 
%and provide the saturation mechanism~
\cite{Kaiser:2001jx,Fritsch:2004nx}.
The nuclear spin-orbit potential, on the other hand, is produced by
the coherent action of scalar and vector mean fields, 
representing the in-medium changes of the quark condensates
\cite{Finelli:2003fk,Finelli:2005ni,Cohen:1994wm,Furnstahl:1997tk,Furnstahl:1999ff,noteonspinorbit}.
The  importance of correlated $2\pi$ exchange 
for hypernuclei  was already pointed out in Ref.~\cite{wei_bro}, and 
recently the  in-medium chiral approach has been extended 
to include the strangeness degree of freedom~\cite{Kaiser:2004fe}. 
%(see Tab.~\ref{actors_table} for 
%the relevant particles involved in the calculation)
The long-range $\Lambda$N interaction arising from kaon and $2\pi$ exchange, 
with a $\Sigma$-hyperon and medium insertions in intermediate states, 
has been explicitly calculated in a controlled expansion in powers of the Fermi momentum $k_f$. In Ref.~\cite{Finelli:2007wm} this approach has been 
applied to finite hypernuclei.
%\begin{wraptable}{l}{0.5\textwidth}
%\centering
%\caption{The main degrees of freedom in the present application of the 
%$SU(3)$ chiral perturbation theory. We show the relevant 
%baryons and mesons with the corresponding masses 
%(averaged) and quantum numbers.}
%\vspace{12pt}
%\begin{tabular}{ccccc}
%~         & Mass (MeV)        & I & J & P \\
%\hline
%N         &  939       &   $1/2$ &      $1/2$       &   $+$  \\
%$\Delta$  &  1232      &   $3/2$ &      $3/2$       &   $+$  \\
%$\Lambda$ &  1116      &   $0$   &      $1/2$       &   $+$  \\
%$\Sigma$  &  1193      &   $1$   &      $1/2$       &   $+$  \\
%$K$       &  494       &   $1/2$ &      $0$         &   $-$  \\
%$\pi$     &  135       &   $1$   &      $0$         &   $-$  \\
%\hline
%\end{tabular}
%\label{actors_table}
%\end{wraptable}
%
%%%%%%%%%%%%%%%%%%%%%%%%%%%%%%%%%%%%%%%%%%%%%%%%%%%%%%%%%%%
%%%%%%%%%%%%%%%%%%%%%%%%%%%%%%%%%%%%%%%%%%%%%%%%%%%%%%%%%%%

\subsection{The model}
\label{model}

In order to describe hypernuclei we extend  
the relativistic nuclear energy density functional 
(see Sec. 2.2 of Ref.~\cite{Finelli:2005ni}),
by adding the hyperon contribution:
\begin{equation}
%\mathcal{L} = \mathcal{L}^N + \mathcal{L}^\Lambda  \;.
E_0^{~}[{\rho}] = E_0^N[{\rho}] + E_0^\Lambda[{\rho}] \;,
\end{equation}
where $E_0^N[{\rho}]$ describes the core of protons 
and neutrons (cf. Eq.~(12) in
Ref.~\cite{Finelli:2005ni}), and 
$E_0^\Lambda[{\rho}]$ is the leading-order term 
representing the single $\Lambda$-hyperon,
decomposed in free and interaction parts:
\begin{equation}
% \mathcal{L}^\Lambda = \mathcal{L}^{\Lambda}_{\rm free} 
%+ \mathcal{L}^\Lambda_{\rm int}
E_0^\Lambda[{\rho}] = E_{\rm free}^\Lambda[{\rho}] 
+ E_{\rm int}^\Lambda[{\rho}] \;,
\end{equation}
with
\begin{eqnarray}
%\mathcal{L}^\Lambda_{\rm free} & = & \bar{\psi}_\Lambda 
%\left( i\gamma_\mu \partial^\mu -  M_\Lambda \right) \psi_\Lambda \\
%\mathcal{L}^\Lambda_{\rm int}  & = &
%-G^{(\Lambda N)}_S (\hat{\rho}) \left( \bar{\psi} \psi \right) \left(
%\bar{\psi}_\Lambda \psi_\Lambda \right) 
%-G^{(\Lambda N)}_V (\hat{\rho}) \left( \bar{\psi} \gamma_\mu \psi \right) \left(
%\bar{\psi}_\Lambda \gamma^\mu \psi_\Lambda \right) \; .
E_{\rm free}^\Lambda & = & 
\int d^3r \langle \phi_0 | \bar{\psi}_\Lambda [-i 
\bm{\gamma} \cdot \bm{\nabla} + M_\Lambda ]
\psi_\Lambda |\phi_0 \rangle \\
E_{\rm int}^{\Lambda} & = & 
\int d^3r \left\{ \langle \phi_0 | 
G^{\Lambda}_S ({\rho}) \left( \bar{\psi} \psi \right) \left(
\bar{\psi}_\Lambda \psi_\Lambda \right) | \phi_0 \rangle + \right. \nonumber\\
& ~ & \left. \quad \quad \quad \langle \phi_0 | G^{\Lambda}_{V} 
({\rho}) \left( \bar{\psi} \gamma_\mu \psi \right) \left(
\bar{\psi}_\Lambda \gamma^\mu \psi_\Lambda \right) | \phi_0 \rangle
\right\} \; .
\end{eqnarray} 
Here $|\phi_0 \rangle$ denotes the (hypernuclear) ground state.
$E_{\rm free}^{\Lambda}$ is the contribution to the energy from
the free relativistic
hyperon including its rest mass $M_\Lambda$. The interaction term
$E_{\rm int}^{\Lambda}$ includes density-dependent hyperon-nucleon vector
($G^{\Lambda}_V$) and scalar ($G^{\Lambda}_S$) couplings.
They include mean-field contributions from in-medium changes of 
the quark condensates (identified with superscript $(0)$), 
and from in-medium pionic fluctuations governed by 
two-pion exchange processes (with superscript $(\pi)$):
\begin{equation}
G^{\Lambda}_i({\rho}) = G^{\Lambda (0)}_i + G^{\Lambda (\pi)}_i({\rho})
\quad {\rm with} \quad i=S,V \; .
\end{equation}
Minimization of the ground-state energy leads to coupled 
relativistic Kohn-Sham equations for 
the core nucleons and the single $\Lambda$-hyperon. 
Using the notation of Ref.~\cite{Finelli:2005ni}, they read:
\begin{eqnarray}\
\label{dir_eq}
\left[ -i \bm{\gamma} \cdot \bm{\nabla} + M_N + \gamma_0 \left( \Sigma_V +
 \Sigma_R + \tau_3 \Sigma_{TV} \right) + \Sigma_S + \tau_3 \Sigma_{TS} \right] 
\psi_k & = & \epsilon_k \psi_k \\
\label{dir_eq_2}
\left[ -i \bm{\gamma} \cdot \bm{\nabla} + M_\Lambda + \gamma_0 \Sigma_V^\Lambda +
 \Sigma_S^\Lambda \right] \psi_\Lambda & = & \epsilon_\Lambda \psi_\Lambda \; ,
\end{eqnarray}
where $\psi_k$ and $\psi_\Lambda$ denote the single-particle wave functions of 
the nucleon and the $\Lambda$, respectively. 
The single-particle Dirac equations are
solved self-consistently in the ``no-sea'' approximation~\cite{SerotFurnstahl}.
It is important to note that the rearrangement 
self-energy $\Sigma_R$~\cite{Fuchs:1995as} 
is confined to the nucleon sector because all the density dependent couplings are 
polynomials in $k_f$ (and consequently in fractional powers of the baryon density
through the relation $\rho = 2\, k_f^3/(3 \pi^2)$), and there is no hyperon Fermi sea.
The $\Lambda$ self-energies read
\begin{equation}
\Sigma_V^\Lambda  = G^{\Lambda}_V(\rho)\, \rho~ ,~~~~~~~~~~~~~~~~
\Sigma_S^\Lambda  =  G^{\Lambda}_S(\rho)\, \rho_S~  ,
\end{equation}
expressed in terms of the nuclear baryon and scalar 
densities, $\rho$ and $\rho_S$. 

In the following sections we  analyze separately
the different contributions to the density-dependent 
$\Lambda$-nuclear couplings
$G^{\Lambda}_i({\rho})$, arising from the kaon- and 
two-pion-exchange induced $\Lambda$-nucleus 
potential, the condensate background mean-fields, 
and the pionic $\Lambda$-nucleus spin-orbit interaction.

%%%%%%%%%%%%%%%%%%%%%%%%%%%%%%%%%%%%%%%%%%%%%%%%%%%%%%%%%
%%%%%%%%%%%%%%%%%%%%%%%%%%%%%%%%%%%%%%%%%%%%%%%%%%%%%%%%%

\subsection{Kaon- and two-pion exchange induced mean-field}
\label{meanfield}

The density-dependent 
self-energy of a zero-momentum $\Lambda$-hyperon 
in isospin-symmetric nuclear matter has been calculated 
in Ref.~\cite{Kaiser:2004fe}
at two-loop order in the energy density. 
This calculation systematically includes kaon-exchange 
Fock terms (first diagram in Fig.~\ref{figL}), 
and two-pion exchange with a $\Sigma$-hyperon 
and including Pauli blocking effects in intermediate states.
\begin{center}
\vspace{-.5cm}
\begin{SCfigure}[][h]
\includegraphics[scale=.52]{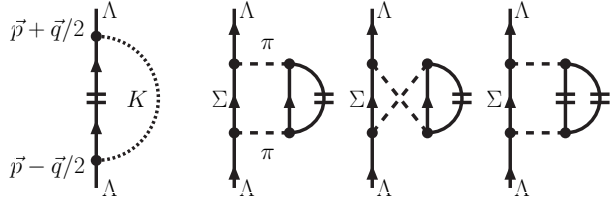}
\caption{\label{figL} One-kaon exchange Fock diagram and two-pion 
exchange Hartree diagrams 
with a $\Sigma$-hyperon in intermediate states. The horizontal double-lines 
represent the filled Fermi sea of nucleons in the in-medium 
propagator: 
$(\gamma\cdot p -M_N)[i(p^2 - M_N^2 + i\epsilon)^{-1} -
2\pi\delta(p^2 - M_N^2)\theta(p_0)\theta(k_f - |\vec{p}\,|)]$~\cite{Kaiser:2004fe}.}
\label{figL}
\end{SCfigure}
\vspace{-.5cm}
\end{center}
The self-energy is translated into a mean field $\Lambda$ 
nuclear potential $U_\Lambda (k_f)$.
A cutoff scale ${\bar{\Lambda}} \simeq 700$ MeV 
(or equivalently, a contact term) represents short-distance
(high momentum) dynamics not resolved at scales characteristic 
for the nucleon Fermi momentum. The value of 
${\bar{\Lambda}}$ is adjusted to reproduce the empirical 
depth of the $\Lambda$ nuclear central potential.

\begin{figure}[h]
\begin{center}
\includegraphics[scale=0.5,angle=0]{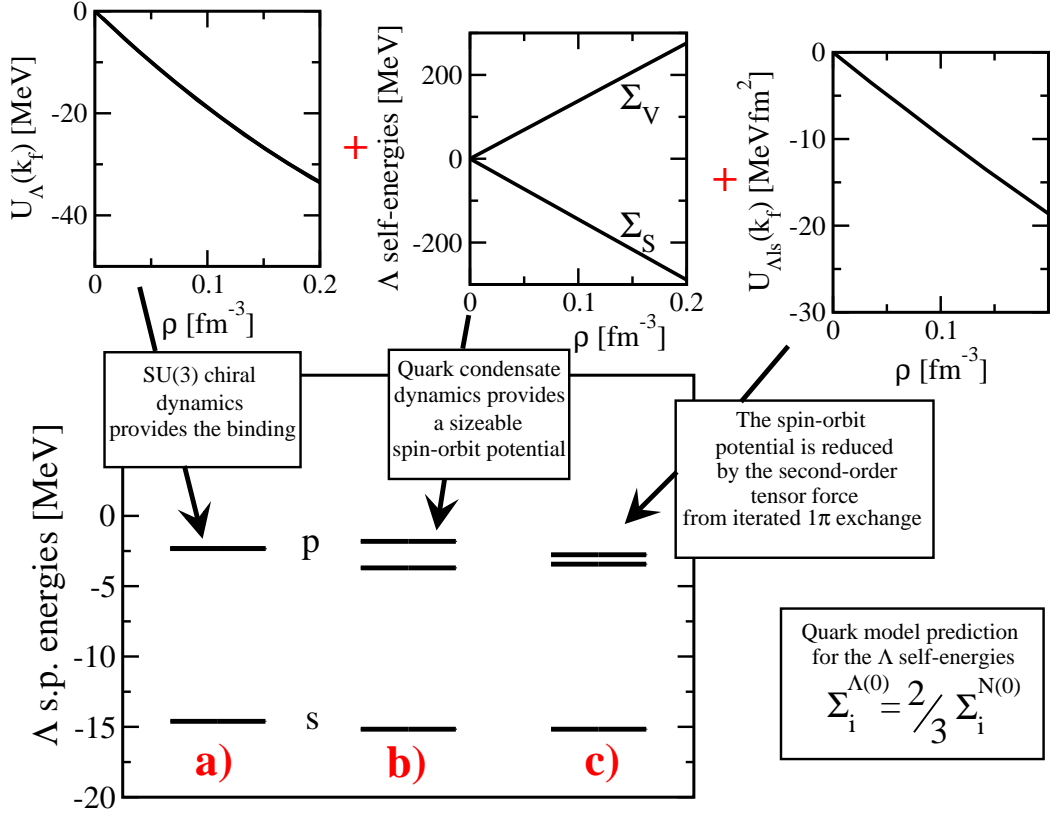}
\vspace{.1cm}
\caption{\label{figC} 
$\Lambda$ single-particle energy levels in $^{17}_{\Lambda}{\rm O}$: 
(a) the contribution of the single-particle potential with
density-dependent coupling strengths 
determined in Ref.~\cite{Kaiser:2004fe} by
including chiral $K$- and $2\pi$-exchange (see Sec.~\ref{meanfield});
(b) the spin-orbit effect from in-medium quark condensates has been included, 
with $\chi = 2/3$ corresponding to the simple quark-model 
prediction~\cite{noblepirnerjenn} (see Sec.~\ref{backfield});
(c) the result of the additional compensating effect 
from the second-order $\Lambda N$ tensor force with intermediate $\Sigma$ 
(see Sec.~\ref{sochiral})).
}
\end{center}
\vspace{-.5cm}
\end{figure}

Following the procedure outlined in  
Appendix A of Ref.~\cite{Finelli:2005ni}, we 
determine the equivalent density dependent 
$\Lambda$ point coupling vertices
$G^{\Lambda (\pi)}_S({\rho})$ and $G^{\Lambda (\pi)}_V({\rho})$.
For the nucleon sector of the energy density functional 
the parameter set FKVW~\cite{Finelli:2005ni} is used.
%: 
%four parameters related to contact
%terms that appear in the ChPT treatment of nuclear matter, one
%parameter for the derivative (surface) term, and two more for
%the strengths of the condensate background scalar and vector
%mean fields.
In Fig.~\ref{figC} (case {\it a}) the $\Lambda$ single-particle 
energy levels are plotted 
for the closed core of 16 nucleons $(8n+8p)$ 
plus a single hyperon 
($^{17}_{\Lambda}{\rm O}$).
At this stage of the calculation the {\it p}-shell spin-orbit
partners are practically degenerate, and the energies of the 
doublets are, by construction, close to their empirical values.
Even the calculated energy of the {\it s}-state is realistic, although 
slightly too deep in comparison with data
($\epsilon_\Lambda^s =-12.42 \pm 0.05$ MeV for 
$^{16}_\Lambda {\rm O}$~\cite{Hashimoto:2006aw}).

Up to this point the in-medium chiral SU(3) dynamics 
(including $K$- and $2\pi$-exchange) provides the necessary 
binding of the system, but no spin-orbit force. As it has already been
shown in Ref.~\cite{Finelli:2005ni}, the inclusion of derivative couplings 
does not remove the degeneracy of the spin-orbit partner states.

%%%%%%%%%%%%%%%%%%%%%%%%%%%%%%%%%%%%%%%%%%%%%%%%%%%%%%%%%%%%%%%%
%%%%%%%%%%%%%%%%%%%%%%%%%%%%%%%%%%%%%%%%%%%%%%%%%%%%%%%%%%%%%%%%

\subsection{Background scalar and vector mean-fields}
\label{backfield}

In contrast to the mean-field induced by kaon- and 
two-pion exchange, the condensate background  
$\Lambda$ self-energies $\Sigma_V^{\Lambda (0)}$ and $\Sigma_S^{\Lambda (0)}$
produce a sizeable spin-orbit potential, 
analogous to the nucleon case~\cite{Finelli:2005ni}.
Finite-density QCD sum rules predict moderate Lorentz
scalar and vector self-energies for the $\Lambda$ hyperon.
Under some reasonable assumptions about the density dependence of 
certain four-quark condensates~\cite{Jin:1993fr}\footnote{
While $\Sigma_V^{\Lambda (0)}$
is rather insensitive to the details of the calculation, 
$\Sigma_S^{\Lambda (0)}$ can only attain realistic values if 
the four-quark condensate
$\langle\overline{q}q\rangle_{\rho_N}^2$ depends weakly on the
nucleon density, 
%({\it i.e.\/}, if $f$ is small) 
and the four-quark condensate
$\langle\overline{q}q\rangle_{\rho_N}\langle\overline{s}s\rangle_{\rho_N}$
has a strong density-dependence~\cite{Jin:1993fr}.
},
one expects a reduction of the corresponding couplings\footnote{
For simplicity the same reduction factor $\chi$ is used both for the scalar and vector
self-energies. In general one could use two parameters $(\chi_S~{\rm and}~\chi_V)$, 
but the difference can be easily absorbed by the cut-off $\bar{\Lambda}$. 
}
\begin{equation}
G^{\Lambda (0)}_{S,V} =  \chi \,G^{(0)}_{S,V} ~,
\end{equation}
by a factor $\chi$,
where $G^{(0)}_V$ and $G^{(0)}_S$ are
the vector and the scalar couplings to nucleons 
arising from in-medium changes
of the quark condensates, $\langle\bar{q}q\rangle$ 
and $\langle q^\dagger q\rangle$.
The values of $G^{(0)}_V$ and $G^{(0)}_S$ have been determined 
by fitting to ground-state properties 
of finite nuclei~\cite{Finelli:2005ni}, and found to be
in good agreement with leading-order QCD sum 
rules estimates~\cite{Cohen:1994wm}.

In Fig.~\ref{figC} (case {\it b}) we plot the $\Lambda$ single-particle 
energy levels calculated with the inclusion of these scalar 
and vector mean-fields, using the quark model prediction 
for the reduction parameter $\chi = 2/3$~\cite{noblepirnerjenn}. 
In this case the {\it p}-shell spin-orbit partners are no longer degenerate.
The calculated spin-orbit splitting is of the order of $\sim 2$ MeV.
However, the choice $\chi = 2/3$ is a rather simplistic estimate.
A detailed  QCD sum rule analysis suggests a reduction to 
$\chi \sim 0.4 - 0.5$~\cite{Cohen:1994wm,Kaiser:2004fe,Jin:1993fr}, and to
even smaller values if corrections from in-medium
condensates of higher dimensions are taken into account.
We note that the $\Lambda$-nuclear 
spin-orbit force is evidently still far too strong 
at this level, just as in the phenomenological 
relativistic "sigma-omega" mean-field models.

\subsection{$\Lambda$-nuclear spin-orbit interaction from 
chiral SU(3) two-pion exchange}
\label{sochiral}

The $\Lambda$-nucleus spin-orbit interaction generated by the
in-medium two-pion exchange $\Lambda$N interaction
has been evaluated in Ref.~\cite{Kaiser:2004fe}. In the spin-dependent 
part of the self-energy of a $\Lambda$ hyperon 
scattering in slightly inhomogeneous nuclear matter 
from initial momentum $\vec{p} - \vec{q}/2$ to final
 momentum $\vec{p} + \vec{q}/2$, one identifies the 
spin-orbit term $\Sigma_{ls}^\Lambda(k_f) = 
{i\over 2}U_{ls}^\Lambda(k_f)\,
\vec{\sigma}\cdot(\vec{q}\times\vec{p}\,)$.
It depends only on known $SU(3)$ axial-vector 
coupling constants and on the mass difference 
between the $\Lambda$ and $\Sigma$. The relevant 
momentum space loop integral is finite, and 
hence model independent in the sense that no 
regularizing cutoff is required. 
The result, 
\[
 U_{ls}^\Lambda(k_f^{(0)})\simeq -15~{\rm MeV~fm}^2 \quad {\rm at}
\quad k_f^{(0)} \simeq 1.36~{\rm fm}^{-1} \; ,
\]
has a sign {\it opposite} to the standard nuclear spin-orbit 
interaction\footnote{Recall that nuclear Skyrme phenomenology gives
$U_{ls}^N(k_f^{(0)}) = 3 W_0 \rho_0/2  \simeq 30$ MeV fm$^2$ for 
the strength of the nucleon spin-orbit potential~\cite{Chabanat:1998}.}. 
This term evidently tends to largely cancel
the spin-orbit potential generated by the scalar 
and vector background mean-fields.
It is important to note that such a ``wrong-sign" 
spin-orbit interaction (generated by the second-order tensor force 
from iterated pion exchange) 
exists also for nucleons \cite{Kaiser:2002jz}.
However, this effect is compensated to a large 
extent by the three-body spin-orbit force involving 
virtual $\Delta(1232)$-isobar excitations \cite{Kaiser:2003ux} (see Fig.~\ref{figD}), 
so that the spin-orbit interaction from the strong 
scalar and vector mean-fields prevails. 
For a $\Lambda$-hyperon, on the other hand, 
the analogous three-body effect does not 
exist, and the cancelation is now between spin-orbit 
terms from the (weaker) background mean-fields and 
the in-medium second-order tensor force from iterated 
pion-exchange with intermediate $\Sigma$. 
The small $\Sigma-\Lambda$ mass splitting, 
$M_\Sigma - M_\Lambda = 77.5$ MeV, plays 
a prominent role in this mechanism.
\begin{center}
\vspace{-.3cm}
\begin{SCfigure}[][h]
\sidecaptionsep
%\begin{center}
$\quad$ \includegraphics[scale=0.65]{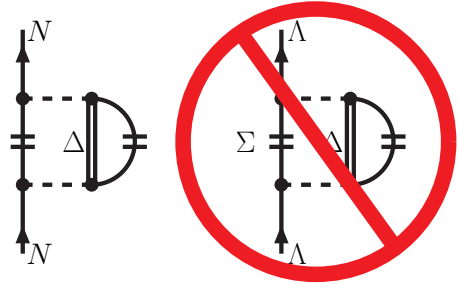}
%\end{center}
\caption{\label{figD}
Three-body diagram of two-pion exchange with virtual 
$\Delta(1232)$-isobar excitation (left). For a nucleon it generates a sizeable 
three-body spin-orbit force of the ``right sign''. The horizontal double-line 
denotes the filled Fermi sea of nucleons. The analogous diagram does not
exist for a $\Lambda$-hyperon (right), simply because replacing the external 
nucleon by a $\Lambda$-hyperon introduces as the intermediate state
on the open baryon line a $\Sigma$-hyperon, for which there is no filled
Fermi sea.}
\end{SCfigure}
\vspace{-.5cm}
\end{center}
In order to estimate the impact of 
this genuine ``wrong-sign" $\Lambda$-nuclear spin-orbit term, we introduce
\begin{equation}
\Delta\mathcal{H}_{ls}^{\Lambda} = 
-i\,{U_{ls}^\Lambda(k_f^{(0)})\over 2r}\,{df(r)\over dr}\,
\vec{\sigma}\cdot(\vec{r}\times\vec{\nabla})~, 
\end{equation}
with the form-factor determined by the normalized 
nuclear density profile $f(r) = \rho(r)/\rho(r=0)$.
The corrections to the $\Lambda$ single particle energies $\epsilon_{\Lambda}$ 
are then evaluated in first-order perturbation theory:
\begin{equation} 
\epsilon_{\Lambda}' = \epsilon_{\Lambda} + 
\langle \phi | \Delta\mathcal{H}_{ls}^{\Lambda} | \phi \rangle \;,
\label{eqnls}
\end{equation}
where $|\phi \rangle$ denotes the self-consistent solution of the
system of Dirac single-baryon equations (\ref{dir_eq})
and (\ref{dir_eq_2}).
In Fig.~\ref{figC} (case {\it c}) one observes that 
the resulting {\it p}-shell single-particle energy levels, 
corrected according to Eq.(\ref{eqnls}), are close to being degenerate. 
The spin-orbit splitting is now strongly reduced, but still rather
too large in comparison with empirical estimates.
This could be a consequence of the possibly too large 
quark-model reduction factor $\chi = 2/3$. We note that within  
the range of values of $\chi$ compatible with QCD sum rules 
estimates, the empirical, almost vanishing 
spin-orbit splitting for the single-$\Lambda$ states can indeed be 
obtained~\cite{Finelli:2007wm}.

In Fig.~\ref{Fig_dep} we plot the $\Lambda$ spin-orbit spacing
$\delta_\Lambda = \epsilon_\Lambda (1p^{-1}_{1/2}) - 
\epsilon_\Lambda (1p^{-1}_{3/2})$ 
as a function of the ratio
$\chi$ between the background mean-fields 
for the $\Lambda$-hyperon and for the nucleon. The
circles denote the spin-orbit splittings produced by the 
scalar and vector background fields alone. Even for 
unnaturally small values of $\chi$, the splitting 
remains systematically too large in comparison with 
empirical estimates. Introducing the model-independent 
spin-orbit contribution from second-order
pion exchange (Eq.(\ref{eqnls})), these values 
are systematically reduced by about 1.3 MeV (triangles). 
For $\chi$ in the range 
$0.4 - 0.5$ determined by the QCD sum rule analysis of 
Refs.~\cite{Cohen:1994wm,Jin:1993fr},  
the small spin-orbit splitting is now reproduced
in agreement with recent empirical values~\cite{Hashimoto:2006aw}.
\begin{center}
\begin{SCfigure}[][h]
\vspace{-.4cm}
\includegraphics[scale=0.3,angle=0]{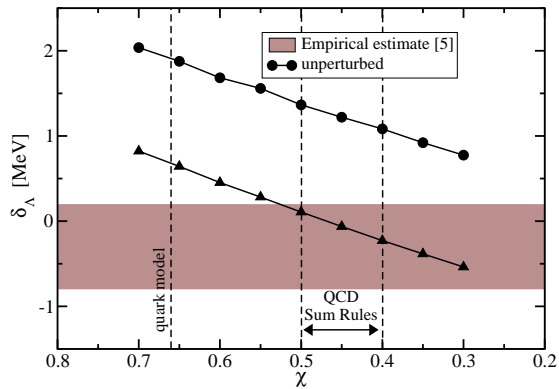}
\caption{\label{Fig_dep} 
Evolution of the spin-orbit splitting
$\delta_\Lambda = \epsilon_\Lambda (1p^{-1}_{1/2}) - 
\epsilon_\Lambda (1p^{-1}_{3/2})$ 
in $^{17}_{\Lambda}{\rm O}$, as a function of the ratio
$\chi$ between the background self-energies of the $\Lambda$  and nucleon. 
The dashed line at $\chi = 2/3$ denotes the simple quark-model value. 
Also indicated is the $\chi$-interval allowed by the QCD sum rule analysis 
of Refs.~\cite{Cohen:1994wm,Kaiser:2004fe,Jin:1993fr}. 
Calculations with (without) the chiral SU(3) spin-orbit correction 
(Eq.~(\ref{eqnls})) are denoted by triangles (circles).
The shaded area represents an estimate of $\delta_\Lambda$ 
($-0.8$ MeV $\le \delta_\Lambda \le 0.2$ MeV) based on the measured 
energy difference $\Delta E (2_1^+ -0_1^+)$ in
$~^{16}_\Lambda$O~\cite{Hashimoto:2006aw}. 
}
\end{SCfigure}
\vspace{-1.cm}
\end{center}

%%%%%%%%%%%%%%%%%%%%%%%%%%%%%%%%%%%%%%%%%%%%%%%%%%%%%%%%%%
%%%%%%%%%%%%%%%%%%%%%%%%%%%%%%%%%%%%%%%%%%%%%%%%%%%%%%%%%%

\section{Conclusions}% from the preprint
The compensating mechanism 
for the spin-orbit interaction of the $\Lambda$-hyperon in nuclear 
matter, suggested in Ref.~\cite{Kaiser:2004fe}, 
successfully explains the very small spin-orbit 
splittings in finite $\Lambda$ hypernuclei. We emphasize 
that this mechanism, driven by the second-order 
pion-exchange tensor force between $\Lambda$ and nucleon, 
with intermediate $\Sigma$-states, is model-independent in 
the sense that it relies only on SU(3) chiral dynamics with 
empirically well determined constants. This 
intermediate-range effect (independent of any 
regularization procedure) counteracts short-distance 
spin-orbit forces. In ordinary nuclei, i.e. without hyperons, 
the corresponding effect is neutralized by three-body 
spin-orbit terms (induced by two-pion exchange with virtual
$\Delta$-isobar excitations). These terms are absent in hypernuclei. 
The theoretical framework based on in-medium chiral dynamics, 
described in this work, will be systematically applied to heavier 
hypernuclei. 

\vspace{.2cm}
\noindent
{\bf Acknowledgements}\\
I would like to thank my collaborators Dario Vretenar, Norbert Kaiser and Wolfram
Weise, and acknowledge discussions with Avraham Gal and Achim Schwenk. 
This work was supported by 
INFN and MURST.
%
 %include also a a very brief explanation of the C_5 term

\end{document}